\documentclass[algorithms,article,accept,moreauthors]{Definitions/mdpi} 

\firstpage{1} 
\pubvolume{18}
\issuenum{8}
\articlenumber{491}
\pubyear{2025}
\copyrightyear{2025}
\externaleditor{Esam El-Araby} 
\datereceived{8 July 2025} 
\daterevised{3 August 2025} 
\dateaccepted{5 August 2025 }
\datepublished{6 August 2025 } 
\hreflink{https://doi.org/10.3390/a18080491} 
\pdfoutput=1 

\usepackage{braket}

\newcommand{\Eq}[1]{Equation~(\ref{#1})}

\newcommand{\Cmbb}{\mathbb{C}}

\makeatletter
\NewDocumentCommand{\textsubstack}{O{}m}
 {
  \textnormal{%
    \fontsize{\f@size}{\fpeval{\f@size*\f@baselineskip/10}}
    \selectfont
    #1
    \begin{tabular}{@{}c@{}}#2\end{tabular}%
  }%
}
\makeatother

\Title{An Early Investigation of the HHL Quantum Linear Solver for Scientific Applications}

\TitleCitation{An Early Investigation of the HHL Quantum Linear Solver for Scientific Applications}

\Author{Muqing Zheng *\orcidA{},
Chenxu Liu \orcidB{}, Samuel Stein \orcidC{}, Xiangyu Li  \orcidD{}, Johannes M\"ulmenst\"adt \orcidE{}, Yousu Chen \orcidF{} \linebreak  and Ang Li *\orcidG{}}

\AuthorNames{Muqing Zheng, Chenxu Liu, Samuel Stein, Xiangyu Li, Johannes M\"ulmenst\"adt, Yousu Chen, and Ang Li}

\isAPAStyle{%
       \AuthorCitation{Zheng, M., Liu, C., Li, X., M\"ulmenst\"adt, J., Chen, Y., \& Li, A.}
         }{%
        \isChicagoStyle{%
        \AuthorCitation{Zheng Muqing, Chenxu Liu, Samuel Stein, Xiangyu Li, Johannes M\"ulmenst\"adt, Yousu Chen, and Ang Li}
        }{
        \AuthorCitation{Zheng, M.; Liu, C.; Li, X.; M\"ulmenst\"adt, J.; Chen, Y.; Li, A.}
        }
}

\address[1]{%
Pacific Northwest National Laboratory, Richland, WA 99354, USA; chenxu.liu@pnnl.gov (C.L.); samuel.stein@pnnl.gov (S.S.); xiangyu.li@pnnl.gov (X.L.); johannes.muelmenstaedt@pnnl.gov (J.M.);  yousu.chen@pnnl.gov (Y.C.)
}

\corres{\hangafter=1 \hangindent=1.05em \hspace{-0.82em} Correspondence: muqing.zheng@pnnl.gov (M.Z.); ang.li@pnnl.gov (A.L.)}

\abstract{In this paper, we explore using the Harrow--Hassidim--Lloyd (HHL) algorithm to address scientific and engineering problems through quantum computing, utilizing the NWQSim simulation package on a high-performance computing platform. Focusing on domains such as power-grid management and climate projection, we demonstrate the correlations of the accuracy of quantum phase estimation, along with various properties of coefficient matrices, on the final solution and quantum resource cost in iterative and non-iterative numerical methods such as the Newton--Raphson method and finite difference method, as well as their impacts on quantum error correction costs using the Microsoft Azure Quantum resource estimator. We summarize the exponential resource cost from quantum phase estimation before and after quantum error correction and illustrate a potential way to reduce the demands on physical qubits. This work lays down a preliminary step for future investigations, urging a closer examination of quantum algorithms' scalability and efficiency in domain applications.}


\begin{document}


\section{Introduction}

Starting with the Deutsch–Jozsa algorithm and Shor's discrete logarithm algorithm~\cite{deutsch, shor1994}, the~potential of quantum computing algorithms has extended beyond merely simulating quantum systems. The~potential speedup of quantum algorithms over their classical counterparts has gathered tremendous attention, including a fundamental demand in science and engineering: solving linear systems. Harrow, Hassidim, and~Lloyd (HHL) first developed a quantum linear solver with an exponential speedup in problem dimensions in~\cite{HHL}. Built upon the exponential speedup of quantum linear system algorithms (QLSAs), many works have explored theoretical quantum advantages in various applications. These fields include portfolio optimization~\cite{rebentrost2018quantum}, machine learning~\cite{HHLsurvey, liu2024towards}, differential equation solving~\cite{liu2021de}, linear optimization~\cite{casares2020quantum, mohammadisiahroudi2022efficient, mohammadisiahroudi2023improvements, wu2023inexact}, and~semi-definite optimization~\cite{augustino2023quantum, mohammadisiahroudi2023quantum}.

However, the~HHL algorithm proposed in~\cite{HHL} has a quadratic dependency on matrix condition number and matrix sparsity, worse than classical linear solvers such as factorization methods and conjugate gradient, where condition number is the product of the norm of the coefficient matrix and the norm of the inverse matrix.
Several works have been proposed to reduce the dependency on the condition number of coefficient matrices and the accuracy of the solution state~\cite{ambainis2012, clader2013preconditioned, childs17, chakraborty18, subasi19, lin22, costa2022, vazquez2022enhancing, jennings2023}. Specifically, based on adiabatic theorems, the~state of the art has a linear or quasi-linear dependency on the condition number and a logarithmic dependency on the inverse of the solution accuracy~\cite{subasi19, lin22, costa2022, jennings2023}.

The HHL algorithm has been demonstrated in experiments to solve linear algebra problems. The~largest linear systems demonstrated on real gate-based quantum machines are up to $4 \times 4$ systems with variants of the HHL algorithm~\cite{yalovetzky2021nisq, Savarsson22, morgan2024enhanced} and an $8 \times 8$ system with the linear solver based on adiabatic quantum computing~\cite{wen2019experimental}. However, testing QLSAs on real quantum devices to demonstrate a quantum advantage still suffers from multiple obstacles, such as the large number of required quantum gates and the high noise level of current quantum devices~\cite{Preskill2018}.

With the current development of quantum hardware and exploration of quantum error correction (QEC) codes, a~large-scale fault-tolerant quantum computer is expected to be demonstrated in the foreseeable future~\cite{krinner2022realizing, google2023suppressing, wang2023fault, mayer2024benchmarking, bluvstein2024logical, da2024demonstration, gupta2024encoding}. QEC codes, such as surface codes, are expected to detect and correct Pauli errors, as~well as any linear combinations of them, provided the errors occur below a certain threshold probability~\cite{nielsen2001quantum}.
Although the gap between algorithm requirements and hardware specifications is shrinking, the~gap still exists, which necessitates the analysis of the resource costs involved~\cite{pascuzzi2022importance}. Resource estimations have been performed for chemistry~\cite{fedorov2022quantum}, Grover's algorithm on the Advanced Encryption Standard~\cite{grassl2016applying}, Shor's discrete logarithm algorithm for the RSA cryptosystem~\cite{gidney2021factor}, and~the computation of elliptic curve discrete logarithms~\cite{roetteler2017quantum}. However, despite this being essential for understanding the disparity between hardware capabilities and practical applications, there is limited work on non-asymptotic resource estimation for QLSAs~\cite{scherer2017concrete}.

In this paper, we focus on resource estimation and experiment with the HHL algorithm on several applications selected from domain science, such as power grid and climate projection. Different from the previous works about asymptotic and non-asymptotic resource analysis~\cite{HHL, ambainis2012, clader2013preconditioned, childs17, scherer2017concrete, chakraborty18, subasi19, lin22, costa2022, vazquez2022enhancing, jennings2023}, we investigate the factors affecting the final accuracy, resource cost, and~fault-tolerant hardware requirements. Our experiments show the effectiveness of the HHL algorithm in scientific applications with a low accuracy in quantum phase estimation. Working with the Microsoft Azure Quantum resource estimator~\cite{beverland2022assessing, van2023using}, we summarize the exponential dependency of quantum resources on the number of clock qubits in HHL circuits and demonstrate a possible method to reduce the demands on physical qubits in fault-tolerant quantum~computing.

This paper is organized as follows: Section~\ref{sec:background} introduces the idea of quantum linear system solvers, with~implementation-related details. Section~\ref{sec:tools} presents the simulator, NWQSim~\cite{nwqsim}, and~the resource estimation tool. Next, we explore the factors of interest in evaluating numerical experiments in Section~\ref{sec:factors} and perform those experiments in Section~\ref{sec:applications}. Finally, we discuss the limitations in Section~\ref{sec:discussion} and conclude the implications of our work on domain science applications in Section~\ref{sec:conclusion}.



\section{Quantum Linear Systems and the Implementation of the~Solver}\label{sec:background}
\unskip

\subsection{Overview of the Harrow--Hassidim--Lloyd (HHL) Algorithm}\label{sec:overview}

Quantum information is encoded into the state of quantum systems. Here, we assume all relevant quantum states can be represented as statevectors. An~$n_d$-qubit statevector $\ket{x} = \sum_{j = 0}^{2^{n_d}-1} \alpha_j \ket{\vec{j}}$ is a normalized complex vector, i.e.,~$\alpha_j \in \Cmbb$ for all $j$ and $\sum_{j = 0}^{2^{n_d}-1} |\alpha_j|^2 = 1$, while $\vec{j} \in \{0,1\}^{n_d}$ is the number $j$ as a binary string. The~set $\{\ket{\vec{j}}\}$ forms the basis set of $\Cmbb^{2^{n_d}}$, referred to as the computational basis. Specifically, $\ket{\vec{j}}$ is the unit vector whose $(j+1)^{\text{th}}$ entry is $1$ and other entries are $0$. The~notation $\bra{\vec{j}}$ is the conjugate transpose of $\ket{\vec{j}}$.

\begin{Definition}[A quantum linear-system problem]\label{def:qls}
  A quantum linear-system problem is to solve a system of linear equations with a normalized solution vector $\ket{x} = A^{-1}\ket{b}/\|A^{-1}\ket{b}\|_2$ where coefficient matrix $A \in \Cmbb^{N \times N}$ is Hermitian and $\ket{x}$ and $\ket{b}$ are both normalized vectors.
\end{Definition}

Start with a classical complex linear system $A \vec{x} = \vec{b}$, $A \in \Cmbb^{N \times N}$, where the right-hand-side (RHS) vector $\vec{b}$ is normalized to obtain $\ket{b} := \vec{b}/\|\vec{b}\|_2$; then, a quantum linear-system problem can be formed with $A$ and $\ket{b}$ if $A$ is Hermitian. Otherwise, a~larger linear system can be constructed as follows 
~\cite{HHL}:\begin{align}
\left[\begin{array}{cc}
\mathbf{0} & A \\
A^\dagger & \mathbf{0}
\end{array}\right]
\left[\begin{array}{c}
\vec{0} \\
\ket{x}
\end{array}\right]
=
\left[\begin{array}{c}
\ket{b} \\
\vec{0}
\end{array}\right],
\label{eq:nonher-hhl}
\end{align}
where $\cdot^\dagger$ is the conjugate transpose. Therefore, we assume coefficient matrices are Hermitian in the rest of this paper. Since the data is encoded into qubits, if~the dimensions of $A$ and $\vec{b}$ are not in the power of 2, $A$ and $b$ must be expanded. 
Suppose there exists a quantum linear system solver that obtains $\ket{x} = A^{-1}\ket{b}/\| A^{-1}\ket{b}\|_2$ from the circuit; then the original solution of the system can be recovered by $\vec{x} = \|b\|_2   \| A^{-1}\ket{b}\|_2  \ket{x}$. While $\|b\|_2$ is known from the previous computation, the~solver needs to provide the value of $ \| A^{-1}\ket{b}\|_2$.

\subsubsection{Mathematical Foundation of~HHL}

Harrow, Hassidim, and~Lloyd first developed the HHL algorithm to solve the quantum linear-system problem~\cite{HHL}. The~fundamental idea behind the HHL algorithm is that the eigenstates of the Hermitian matrix $A$ (noted as $\{\ket{v_j}\}$) form a complete orthonormal basis of $\Cmbb^{N}$ (i.e., $\braket{v_j|v_k} = \delta_{jk}$), and~hence the state $\ket{b}$ can always be decomposed by this basis as $\ket{b}= \sum_{j=0}^{N - 1} b_j \ket{v_j}$. Similarly,
\begin{align}
    \ket{x} &= \frac{A^{-1}\ket{b}}{\|A^{-1}\ket{b}\|_2}  \nonumber\\
            &=\frac{1}{\|A^{-1}\ket{b}\|_2} \sum_{j=0}^{N - 1} \frac{1}{\lambda_j}\ket{v_j}\bra{v_j}  \sum_{j=0}^{N - 1} b_j \ket{v_j} \nonumber\\
            &=  \frac{1}{\sqrt{\sum_{j=0}^{N-1} \frac{|b_j|^2}{\lambda_j^2}}} \sum_{j=0}^{N - 1} \frac{b_j}{\lambda_j}\ket{v_j}. \label{eq:hhl-foundation}
\end{align}

\begin{figure}[H]
    \centering
    \includegraphics[width=0.99\linewidth]{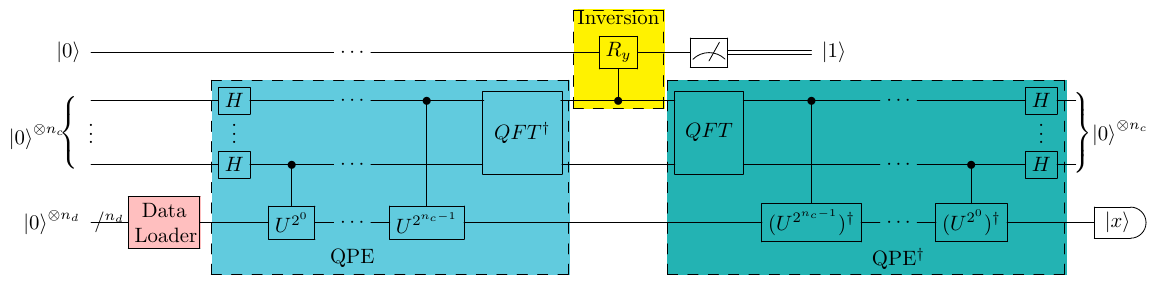}
    \caption{HHL circuit. The~unitary gates in quantum phase estimation (QPE) are $U = e^{itA}$ and $U^{2^j} = e^{i2^jtA}$ where $i^2 =-1$ and $t$ is a scaling factor. The~top qubit is referred to as the ancillary qubit, and~it is the most significant~qubit.}
    \label{fig:hhl-general}
\end{figure}

In other words,  the~HHL algorithm needs a quantum computer to perform eigen-decomposition of $A$ and eigenvalue inversion. Figure~\ref{fig:hhl-general} shows a general description of the circuit that exactly serves the purpose, with~an additional $n_d$-qubit data-loading block to load $\ket{b}$ into the quantum computer and $n_d = \lceil \log(N) \rceil$. 

\vspace{-3pt}

\subsubsection{Quantum Phase~Estimation}

The eigen-decomposition requires a subroutine called quantum phase estimation (QPE), as~illustrated in the sky blue part of Figure~\ref{fig:hhl-general}. Given a unitary matrix $U$ has an eigenstate $\ket{v_j}$ with eigenvalue $e^{2\pi i \theta_j}$, QPE is a quantum algorithm to solve the phase of the eigenvalue ($\theta_j$) \cite{nielsen2001quantum}. After~executing the QPE algorithm, the binary representation of the phase angle $\theta$ is stored in a qubit state. The~qubits carrying the phase information are named ``clock qubits''.  In~the HHL algorithm, if~$\ket{v_j}$ is an eigenstate of a Hermitian matrix $A$ with eigenvalue $\lambda_j$, by~constructing a unitary matrix $U = e^{itA}$ with a scale factor $t$, the~state $\ket{v_j}$ becomes an eigenstate of $U$ with eigenvalue $e^{i t \lambda_j}$. Therefore, the~eigenvalue $\lambda_j$ can be estimated using the QPE~algorithm. 

Suppose we have access to the gate $U$; then it is clear that $U^l \ket{v_j} = e^{2\pi i \theta_j l} \ket{v_j}$ for some positive integer $l$. QPE requires a submodule called quantum Fourier transform (QFT). QFT maps\begin{align*}
    QFT \ket{\vec{j}} &= \frac{1}{\sqrt{2^{n_c}}} \sum_{k = 0}^{2^{n_c} - 1} \omega^{jk} \ket{\vec{k}} \\
    &= \frac{1}{\sqrt{2^{n_c}}}  \left(\ket{0}+e^{2\pi i 0.j_{n_c}}\ket{1}\right)\left(\ket{0}+e^{2\pi i 0.j_{n_c-1}j_{n_c}}\ket{1}\right) \cdots \left(\ket{0}+e^{2\pi i 0.j_1j_2\dots j_{n_c}}\ket{1}\right) 
\end{align*}
where $\vec{j}$ has $n_c$ bits and $\omega = e^{2\pi i /(2^{n_c})}$. Intuitively, $\vec{j}$ can be considered as binary number $\vec{j} = j_1j_2\dots j_{n_c}$ such that $j_{l} \in \{0,1\}$ and QFT transforms this binary number from a state to the phases of bases in different accuracy. So, on~the contrary, if~we apply the inverse of the QFT operator, denoted by $QFT^{\dagger}$, the~phase value becomes a state, and~we can measure the state to obtain the phase value in the binary~representation. 

To summarize the process of a standalone QPE routine, we have\begin{align*}
    \ket{0}^{\otimes n_c} \ket{v_j} &\xrightarrow{H^{\otimes n_c}} \frac{1}{\sqrt{2^{n_c}}}\sum_{k=0}^{2^{n_c}-1} \ket{\vec{k}} \ket{v_j} \\
    &\xrightarrow{CU \text{ sequence}}  \frac{1}{\sqrt{2^{n_c}}}\sum_{k=0}^{2^{n_c}-1} e^{2\pi i \theta_j k} \ket{\vec{k}} \ket{v_j} \\
    &\xrightarrow{QFT^{\dagger}} \ket{\tilde{\theta}_j} \ket{v_j}
\end{align*}
where the $CU \text{ sequence}$ is the controlled-$U$ sequence in the sky blue part of Figure~\ref{fig:hhl-general} and $\tilde{\theta}_j = \theta_j$ if $\theta_j$ can be perfectly represented in $n_c$ bits; otherwise, $\tilde{\theta}_j $ is an estimation of $\theta_j$ in a finite accuracy. In~other words, the~number of clock qubits, $n_c$, governs the accuracy of the estimated eigenvalue in QPE. To~understand more details about QFT and QPE, we direct the interested reader to~\cite{nielsen2001quantum,zaman2023step}.

Note that, without~circuit optimization, the~increase in $n_c$ will exponentially increase the gate counts in HHL circuits. Recall the HHL circuit in Figure~\ref{fig:hhl-general}: an~extra clock qubit leads to an extra controlled $U^{2^{n_c - 1}}$ and an extra controlled inverse $U^{2^{n_c - 1}}$ in the HHL circuit, where $U = e^{itA}$ and $A$ is the coefficient matrix in a linear system. Generally, we should not explicitly compute the matrix $U^{2^{n_c - 1}}$, but~apply gate $U$ for $2^{n_c - 1}$ times in the circuit. Then, the~QPE part of the circuit contains $\sum_{j = 0}^{n_c-1} 2^j = 2^{n_c}-1$ number of $U$, and~when there are $n_c + 1$ clock qubits, an~extra $2^{n_c}$ number of $U$ is added into the QPE, which almost doubles the number of gates $U$ in QPE. The~same situation happens on the inverse QPE part of the HHL circuit.

\subsubsection{State Evolution in~HHL}

In general, the~evolution of states in the HHL circuit is\begin{align*}
    \ket{0}\ket{0}^{\otimes n_c}\ket{0^{n_d}} &\xrightarrow{\text{Data loading}} \ket{0}\ket{0}^{\otimes n_c}\ket{b} \\
                                      &\xrightarrow{\text{QPE}} \sum_{j=0}^{2^{n_{d}}-1} b_j \ket{0}\ket{\tilde{\lambda}_j}\ket{v_j}\\
                                      &\xrightarrow{\textsubstack{Eigenvalue\\Inversion}} \sum_{j=0}^{2^{n_d}-1} b_j \left(\sqrt{1 - \frac{C^2}{\tilde{\lambda}_j^2}}\ket{0} + \frac{C}{\tilde{\lambda}_j}\ket{1}\right)\ket{\tilde{\lambda}_j} \ket{v_j} \\
                                      &\xrightarrow{\textsubstack{Measure the ancillary\\only keep $\ket{1}$}} D \sum_{j=0}^{2^{n_d}-1} \frac{b_j}{\tilde{\lambda}_j}\ket{1}\ket{\tilde{\lambda}_j} \ket{v_j} \\
                                      &\xrightarrow{\text{QPE}^\dagger} D \ket{1}\ket{0^{n_c}} \sum_{j=0}^{2^{n_d}-1} \frac{b_j}{\tilde{\lambda}_j} \ket{v_j} \approx  D \ket{1}\ket{0^{n_c}} \ket{x}
\end{align*}
where $\tilde{\lambda}_j \approx \lambda_j$ is the eigenvalue of $A$ with a finite accuracy estimated in QPE, and~$C$ and $D$ are both constant normalization factors
\begin{align*}
    D = \frac{C}{\sqrt{\sum_{j=0}^{2^{n_d}-1} C^2 \frac{|b_j|^2}{\tilde{\lambda}_j^2}}}  \approx  \frac{1}{\| A^{-1}\ket{b}\|_2}.
\end{align*}
Thus, the~norm $\| A^{-1}\ket{b}\|_2$ can be estimated by the probability of measuring the ancillary qubit in state $\ket{1}$, i.e.,~$\| A^{-1}\ket{b}\|_2 \approx \sqrt{\text{Pr}(\text{Measure 1 in ancillary})}$. This value can be obtained without extra cost as we need to run the circuit multiple times to get $\ket{x}$ or $\bra{x}M\ket{x}$ for some observable $M$. The~overall runtime complexity of HHL algorithm is $\tilde{O}(\log(N)s^2 \kappa^2/\epsilon)$, where $s$ is the sparsity of $A$, $\kappa = \|A\|\|A^{-1}\|$ is the condition number of $A$, and~$\epsilon$ is the final additive error of the solution defined by the ideal state $\ket{x}$ and the result from HHL $\ket{x_{HHL}}$ through $\|\ket{x} - \ket{x_{HHL}}\| \leq \epsilon$ \cite{HHL}.

\subsubsection{Quantum-Classical Data Exchange in~HHL}\label{sec:dataio}

There are two major input models for encoding both matrix $A$ (or $e^{itA}$) and vector $\ket{b}$ into a quantum computer. One is the sparse-access model, used in the HHL algorithm~\cite{HHL}. The sparse-access model is a quantum version of classical sparse matrix computation, and~we assume access to unitaries that calculate the index of the $l^{\text{th}}$ non-zero
 element of the $k^{\text{th}}$ row of a matrix $A$ when given $(k, l)$ as input. A~different input model, now known as the quantum operator input model, is from Low and Chuang~\cite{low2019hamiltonian}. This method is based on the block-encoding of $A$ to allow efficient access to entry values. Its circuit implementation can be found in~\cite{camps2022fable, camps2024explicit}. Meanwhile, this encoding scheme can also be achieved using quantum random access memory~\cite{giovannetti2008architectures, kerenidis2016quantum, kerenidis2020quantum, liu2023quantum}. It requires the complexity $O\left(\text{polylog}\left(N/\epsilon_{BE}\right)\right)$ for realizing an $\epsilon_{BE}$-approximate block-encoding of $A \in \mathbb{C}^{N \times N}$ with quantum random access memory~\cite{kerenidis2016quantum}.

\begin{Definition}[The block-encoding of a matrix]\label{def:block-encoding}
The block-encoding of a matrix $A \in \mathbb{C}^{N \times N}$ is a  unitary operator $U$ such that
    \begin{align*}
        U = \begin{bmatrix}
            A/a &\cdot \\ \cdot &\cdot 
        \end{bmatrix}
    \end{align*}
where $a \geq \|A\|$ is a normalizing constant. In~other words, $U$ and $A$ satisfy, for~some constant $a$ and $n$,
\begin{align*}
    a \left(\bra{0}^{\otimes n} \otimes I_N  \right) U \left( I_N \otimes \ket{0}^{\otimes n}  \right) = A
\end{align*}
where $I_N \in  \mathbb{R}^{N \times N}$ is the identity matrix.
\end{Definition}

To construct the matrix exponential $e^{itA}$ in QPE via block-encoding-based methods, block-encoding first embeds the coefficient matrix $A$ into a larger unitary matrix, and then uses Quantum Signal Processing to achieve the Hamiltonian simulation $e^{itA}$~\cite{qsp}. However, limited by the size of tested matrices in our numerical experiments, block-encoding alone already does not show advantages over Quantum Shannon Decomposition (QSD), a~method that directly decomposes $e^{itA}$ into a sequence of basis gates, in~terms of gate counts and the number of encoding qubits. We illustrate this in Table~\ref{tab:qsd-block}. Note that the coefficient matrix used in Section~\ref{sec:power-grid} is non-Hermitian, so its conversion to a Hermitian matrix using Equation~\eqref{eq:nonher-hhl} brings a disadvantage to QSD in terms of the number of~gates.

\begin{table}[H]
\small
\caption{Resource costs of using FABLE \cite{camps2022fable} to block-encode $A$ and QSD to synthesize $e^{2\pi i A}.$ \label{tab:qsd-block}}

\begin{tabularx}{\textwidth}{c c CCCC} \toprule 
\textbf{Linear System}                                   &\textbf{Problem Dimension}               &\textbf{Method}    &\# \textbf{of Qubits}   & \textbf{\# of} \boldmath{$CX$}  & \textbf{\# of} \boldmath{$U_3$$^1$} \\ \midrule
\multirow{2}{*}{Section
 \ref{sec:power-grid} Iter. 1} & \multirow{2}{*}{$5 \times 5$}   & QSD      & 4             & 100  & 208  \\
                                                &                                 & FABLE    & 7             & 73   & 70  \\ \midrule
\multirow{2}{*}{Section \ref{sec:heat} Three-point}       & \multirow{2}{*}{$9 \times 9$}   & QSD      & 4             & 100  & 208  \\
                                                &                                 & FABLE    & 9             & 268  & 264  \\ \midrule
\multirow{2}{*}{Section \ref{sec:heat} Five-point}       & \multirow{2}{*}{$25 \times 25$} & QSD      & 5             & 444  & 904  \\
                                                &                                 & FABLE    & 11            & 1039 & 1034 \\ 
\bottomrule
\end{tabularx}

\noindent{\footnotesize{\textsuperscript{1} $U_3$ gate is the 1-qubit rotation gates with 3 Euler angles.}}

\end{table}

On the other hand, efficiency in reading $\ket{x}$ could be a potential threat to quantum speedup. The~current state-of-the-art quantum state tomography algorithm is from Apeldoorn~et~al. \cite{van2023quantum}. For~a state $\ket{\psi} = \sum_{j = 0}^{N-1} \beta_j \ket{\vec{j}} \in \mathbb{C}^{N}$, with~probability $1-\delta$, the~pure-state tomography in~\cite{van2023quantum} requires $O\left(  \sqrt{N}/\sigma \cdot  \log\left(N/\delta\right)  \right)$ queries to the unitary oracle that prepares $\ket{\psi}$ from $\ket{00\dots 0}$ to output a vector $\vec{\beta}_{est} \in \mathbb{R}^N$ such that $\|\Re(\vec{\beta}) - \vec{\beta}_{est}\|_{\infty} \leq \sigma$. The~same routine can be applied to $i \ket{\psi}$ to estimate the imaginary~part.

\subsection{Implementation of the Circuit~Generation}

In all experiments in this paper, the~code for the HHL circuit generation comes from a Qiskit-based open-sourced package~\cite{anedumla_quantum_linear_solvers}, which only produces the essential parts of the HHL circuit as colored in Figure~\ref{fig:hhl-general}. We made slight modifications to accommodate the changes in Qiskit 0.46. The~state preparation for $\ket{b}$ uses the algorithm in~\cite{iten2016quantum} that decomposes an arbitrary isometry into the optimized number of single-qubit and CNOT gates, where isometry refers to the inner-product-preserving transformation that maps between two Hilbert spaces; i.e.,~the state preparation is a special case of isometries. {To construct the unitary operator $e^{itA}$ in the QPE stage, the~code directly uses Quantum Shannon Decomposition to synthesize the matrix exponential in the circuit.}


\section{Simulator and Resource Estimation~Tool}
\label{sec:tools}

\begin{figure}[H]
    
    \includegraphics[width=0.6\linewidth]{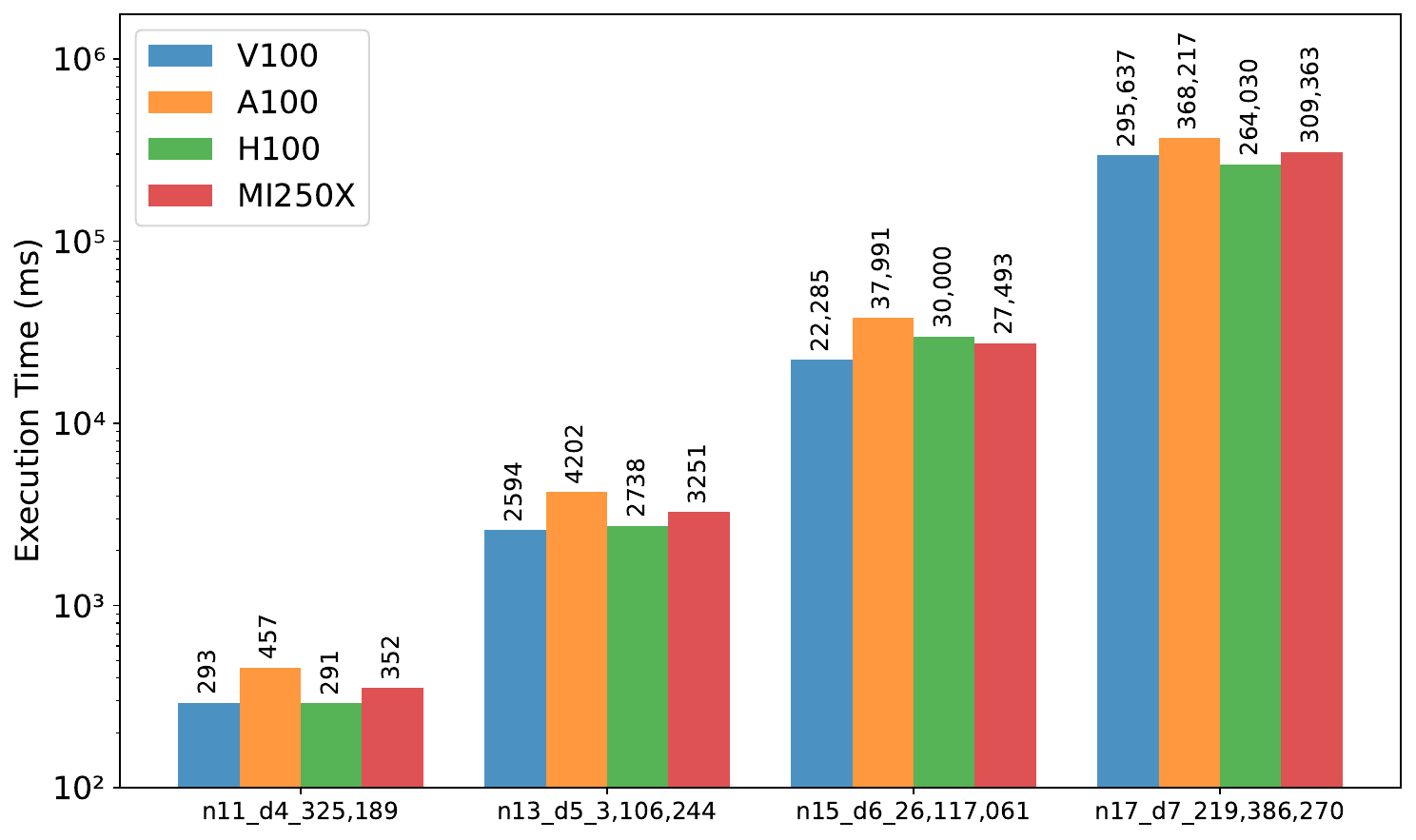}
    \caption{NWQSim performance on different GPUs. The~testing HHL circuits use randomly generated sparse matrices and random RHS vectors. The~three numbers in the name of each testing circuit are the number of qubits in the circuit, the~number of qubits for data loading, and~the total number of gates in the circuit, respectively.}
    \label{fig:gpu-simu-randhhl}
\end{figure}

The statevector simulator carries the simulations in the experiments, SV-Sim~\cite{li2021sv}, in~Northwest Quantum Circuit Simulation Environment (NWQSim)(V2.5.0) \cite{nwqsim}. As shown in ~\cite{li2021sv}, compared to simulators in Aer from Qiskit~\cite{qiskit} and qsim from Cirq~\cite{cirq}, NWQSim provides specialized computation for a wide range of supported basis gates and architectures of CPUs and GPUs, such as gate fusion. In~Table~\ref{tab:hhl-random} and later in Section~\ref{sec:applications}, we demonstrate that the gate fusion strategy in NWQSim can reduce about $80\%$ of gates in the circuits without sacrificing error rates. On~the other hand, NWQSim utilizes a communication model called ``PGAS-based SHMEM'' that significantly reduces communication latency for intra-node CPUs/GPUs and inter-node CPU/GPU clusters. In~this case, SV-Sim has an exceptional performance over other simulators in deep-circuit simulation~\cite{li2021sv}. Figure~\ref{fig:gpu-simu-randhhl} shows the running time of the HHL circuit in the size of 11 qubits to 17 qubits on SV-Sim on four different~GPUs.

\begin{table}[H]
\centering
\caption{HHL circuit properties for four random~examples.\label{tab:hhl-random}}
\begin{tabularx}{\textwidth}{CCCC c C} 
\toprule
\multicolumn{2}{c}{\textbf{No.  of Qubits}} & \multicolumn{1}{c}{\multirow{2}{*}{\textbf{CX Gates}}} & \multicolumn{1}{c}{\multirow{2}{*}{\textbf{Depth}}} & \multicolumn{2}{c}{\textbf{Total No. of Gates}} \\
\boldmath{$n_d$}               & \boldmath{$n_c$}              & \multicolumn{1}{c}{}                          & \multicolumn{1}{c}{}                       & \textbf{Before Fusion}      & \textbf{After Fusion}      \\ \midrule
4                & 6              & 116,535                                        & 248,084                                     & 325,189             & 70,804             \\
5                & 7              & 1,111,178                                       & 2,373,842                                    & 3,106,244            & 665,921            \\
6                & 8              & 9,335,345                                       & 19,969,964                                   & 26,117,061           & 5,557,777           \\
7                & 9              & 78,420,632                                      & 167,816,254                                  & 219,386,270          & 46,631,320        \\ \bottomrule  
\end{tabularx}
\end{table}
\unskip

The resource estimator in~\cite{beverland2022assessing, van2023using} from Microsoft Azure Quantum establishes a systematic framework to access and model the resources necessary for implementing quantum algorithms on a user-specified fault-tolerant scenario. This tool enables detailed estimation of various computational resources, such as the number of physical qubits, the~runtime, and~other QEC-related properties to achieve a quantum advantage for certain applications. Specifically, the~tool accepts a wide range of qubit and quantum error correction code specifications and an error budget that allows different error rates to simulate a described fault-tolerant~environment.

The tool is compatible with circuits generated from a high-level quantum computing language or package, including Qiskit and Q\#. After~a circuit is given, the~input is compiled into Quantum Intermediate Representation through a unified processing program, and~the estimator can examine the code and record qubit allocation, qubit release, gate operation, and~measurement operation. Then, logical-level resources are estimated and used to compute the required physical-level resources further. The~tool returns a thorough report on resources demanded to perform the given algorithm on fault-tolerant quantum computers, including the explanation and related mathematical equations of those estimates. A~selected list of estimates is described in Section~\ref{sec:factors}, and~their values in conducted experiments are displayed in Section~\ref{sec:applications}.

\section{Factors of~Interest}\label{sec:factors}

As we focus on the linear system in scientific applications instead of random systems for benchmarking, we have less control over the specific values of matrix properties like condition numbers. Our interest is more on the number of clock qubits $n_c$ in the HHL circuit, which controls the accuracy of estimated eigenvalues.
The error in eigenvalue estimation affects the solution of the linear system through Equation~\eqref{eq:hhl-foundation}.
From~\cite{nielsen2001quantum}, to~obtain an eigenvalue with $2^{-b}$ accuracy with at least $1-p_{QPE,fail}$ success probability using QPE, we~need
\begin{align*}
    n_c = b + \left\lceil \log_2\left(2 + \frac{1}{2p_{QPE,fail}}\right) \right\rceil.
\end{align*}
In the Qiskit-based HHL implementation that we used~\cite{anedumla_quantum_linear_solvers}, it is suggested that
\begin{align}
    n_c = \max \left( n_d+1,  \left\lceil \log_2(\kappa + 1)   \right\rceil  \right) + \mathbb{I}_{nv} \label{eq:qishhlnc}
\end{align}
where $\mathbb{I}_{nv} = 1$ if the coefficient matrix has a negative eigenvalue; otherwise, it is $0$.
In this paper, we will adjust $n_c$ to illustrate the influence of the QPE resources on the HHL circuit's total cost and the algorithm's accuracy in domain~applications.

When discussing resource estimation under a fault-tolerant setting, our primary concerns are the estimated runtime, the~number of physical qubits, and~extra resources required from the QEC code. We adopt a distance-7 surface code that encodes 98 physical qubits into a single logical qubit. The~theoretical logical qubit error rate is $3~\times~10^{-10}$, and~the error correction threshold is 0.01. The Azure Quantum resource estimator provides several qubit parameter sets to simulate different qubit properties. The~preset qubit settings we used in this paper are $(\text{ns},~10^{-4})$ and $(\text{$\upmu$s},~10^{-4})$ from~\cite{beverland2022assessing}, where the former is close to the specifications of superconducting transmon qubits or spin qubits, and~the latter is more relevant for trapped-ion qubits~\cite{beverland2022assessing}. A~list of detailed configurations of qubit parameter set $(\text{ns},~10^{-4})$ and $(\text{$\upmu$s},~10^{-4})$ is in Table~\ref{tab:er-qubit-config}. We enforce 2-D nearest-neighbor connectivity of the qubits to simulate the connectivity constraint on real quantum computers. So we also demonstrate the changes in some factors before this constraint is enforced (``pre-layout'') and after this constraint is enforced (``after layout'').

\begin{table}[H]
    \centering
    \caption{Qubit parameter~configurations.\label{tab:er-qubit-config}}
    \begin{tabularx}{\textwidth}{RRCC} \toprule
    \multicolumn{2}{r}{\textbf{Qubit Parameter Set}}       & \boldmath{$(\text{ns},~10^{-4})$} & \boldmath{$(\text{$\upmu$s},~10^{-4})$} \\ \midrule
    \multirow{4}{*}{Operation Time} & Measurement       & 100 ns      & 100 $\upmu$s
 \\
                                    & Single-qubit gate & 50 ns       & 100 $\upmu$s      \\
                                    & Two-qubit gate    & 50 ns       & 100 $\upmu$s      \\
                                    & $T$ gate            & 50 ns       & 100 $\upmu$s      \\ \midrule
    \multirow{4}{*}{Error rate}     & Measurement       & $10^{-4}$      & $10^{-4}$      \\
                                    & Single-qubit gate & $10^{-4}$      & $10^{-4}$      \\
                                    & Two-qubit gate    & $10^{-4}$      & $10^{-4}$      \\
                                    & $T$ gate            & $10^{-4}$      & $10^{-6}$       \\ \bottomrule
    \end{tabularx}
\end{table}

Another important tunable parameter is the overall allowed errors for the algorithm, namely error budget. Its parameter value is equally divided into three~parts:
\begin{itemize}
    \item Logical error probability:
the probability of at least one logical error;
    \item T-distillation error probability: the probability of at least one faulty T-distillation;
    \item Rotation synthesis error probability: the probability of at least one failed rotation synthesis.
\end{itemize}

There are also specific breakdowns in the resource required by QEC that are of interest~\cite{beverland2022assessing, van2023using}. We list them in Table~\ref{tab:factors-qec}.

\begin{table}[H]
    
    \caption{Factors of interest in~QEC.}
    \label{tab:factors-qec}
    \footnotesize
    
\begin{adjustwidth}{-\extralength}{0cm}
\begin{tabularx}{\fulllength}{>{\hsize=5.8cm}L>{\hsize=12cm}L}  
    \toprule 
    \textbf{Factors of Interest}     &  \textbf{Description}\\ \midrule
     \multirow{3}{*}{Number of logical qubits pre- and after layout}   &   Under the nearest-neighbor constraint, extra logical qubits could be required to satisfy the \\  
       & connectivity needed in the algorithm (circuit); the relation is $n_\text{after} = 2n_\text{alg} + \left\lceil \sqrt{8 n_\text{alg}} \right\rceil + 1$, where\\
       & $n_\text{alg}$ is the number of logical qubits pre-layout and $n_\text{after}$ is the number of qubits after layout\\ \midrule
     \multirow{2}{*}{Number of physical qubits for the algorithm}   & The product of the number of logical qubits after layout and the number of physical qubits needed\\
        & to encode one logical qubit  \\  \midrule
      Number of physical qubits for $T$ factories  & $T$ factories produce $T$ states to implement non-Clifford operations in a circuit \\  \midrule
      \multirow{2}{*}{Number of physical qubits} & The sum of the number of physical qubits for the algorithm and the number of physical qubits \\ 
      &  for $T$ factories\\ \midrule
      \multirow{2}{*}{Number of $T$ states}  & The estimator requires 1 $T$ state for each of the $T$ gates in a circuit, 4 $T$ states for each of\\
      & the $CCZ$ and $CCiX$ gates, and~18 $T$ states for each of the arbitrary single-qubit rotation gates \\  \midrule
      \multirow{2}{*}{Number of $T$ factories}  & Determined from algorithm runtime, $T$ state per $T$ factory, the~number of $T$ states, and~$T$ factor  \\ 
      &  duration through the equation $\left\lceil \dfrac{T \text{ state} \cdot T \text{ factor duration}}{T \text{ state per $T$ factory} \cdot \text{algorithm runtime}} \right\rceil$ \\ \midrule
      Number of logical cycles for the algorithm  &  The logical depth of the algorithm \\   \midrule
      Min. logical qubit error rate required to run & \multirow{2}{*}{$\dfrac{\text{logical error probability}}{\text{Number of logical qubits}\cdot \text{Number of logical cycles}}$} \\
      the algorithm within the error budget & \\ \midrule
      Min. $T$ state error rate required for  distilled & \multirow{2}{*}{$\dfrac{\text{$T$ distillation error probability}}{\text{ Total number of $T$ states}}$}\\ 
     $T$ states & \\ \bottomrule
    \end{tabularx}
\end{adjustwidth}
    
\end{table}


\section{Scientific Applications and~Evaluation}\label{sec:applications}

This section examines the utilization of the HHL algorithm in the fields of power grids and climate projection. We evaluate the performance of HHL in terms of solution accuracy, resource cost, and~influence on convergence speed for applicable~problems. 

In addition to the hardware specifications in Section~\ref{sec:tools}, all resource estimator jobs are run on the Azure Quantum cloud server. Due to the limitation on the cloud service usage, we cannot examine some of the deepest circuits in this section with the resource estimator, and~all evaluated circuits are transpiled.

with respect to a given basis gate set from the estimator using the transpiler in Qiskit. The~optimization level of the transpiler is set to level 2. The Qiskit version is 0.46. The
Azure Quantum version is 0.30.0. The MATPOWER version is 7.1.

\subsection{Power Flow Problem in Power~Grid}\label{sec:power-grid}

The use of quantum algorithms has drawn much attention in recent research on power system applications, especially the areas where quantum linear system solver can be deployed, including power flow, contingency analysis, state estimation, and~transient simulation~\mbox{\cite{feng2021quantum, chen2022computing, zhou2022quantum, golestan2023quantum, chen2023coming, jing2024hhl}}. The~specific problem type we illustrated in this section is an alternating current power flow~problem. 

The power flow equations are essential to analyzing the steady-state behavior of power systems by describing the relationship between bus voltages (magnitude and phase angles), currents, and~power injections in a power system. The~basic power flow equations are as follows:
\begin{align*}
P_k &= \sum_{j=1}^{n} \left( |V_k| |V_j| \text{Re}(Y_{kj}^*) \cos(\theta_{kj}) + |V_k| |V_j| \text{Im}(Y_{kj}^*) \sin(\theta_{kj}) \right) \\
Q_k &= \sum_{j=1}^{n} \left( |V_k| |V_j| \text{Re}(Y_{kj}^*) \sin(\theta_{kj}) - |V_k| |V_j| \text{Im}(Y_{kj}^*) \cos(\theta_{kj}) \right)
\end{align*}
where
\begin{itemize}
    \item $P_k$: real power injection at bus $k$.
    \item $Q_k$: reactive power injection at bus $k$.
    \item $|V_k|$: voltage magnitude at bus $k$.
    \item $\theta_{kj}$: phase angle difference between bus $k$ and bus $j$.
    \item $Y_{kj}$: admittance between bus $k$ and bus $j$.
\end{itemize}

For a power flow problem with $B$ buses and $G$ generators, there are $2(B-1) - (G-1)$ unknowns representing voltage magnitudes, $|V_k|$, and~phase angles, $\theta_k$, for~load buses and voltage phase angles for generator buses. With~the knowledge of the admittance matrix of the system that represents the nodal admittance of the buses, we can use the Newton--Raphson (N-R) method to solve power flow equation iteratively: after an initial guess for the voltages at all buses, in~each N-R iteration, we solve
\begin{align}
    \begin{bmatrix}
        \frac{\partial \Delta \vec{P}}{\partial \vec{\theta}}  &  \frac{\partial \Delta \vec{P}}{\partial |\vec{V}|} \\
        \frac{\partial \Delta \vec{Q}}{\partial \vec{\theta}}  &  \frac{\partial \Delta \vec{Q}}{\partial |\vec{V}|} \\
    \end{bmatrix}
    \begin{bmatrix}
        \Delta \vec{\theta} \\ \Delta |\vec{V}|
    \end{bmatrix}
    = -
    \begin{bmatrix}
        \Delta \vec{P} \\ \Delta \vec{Q}
    \end{bmatrix}
    \label{eq:power-nr}
\end{align}
where $\Delta P_k$ and $\Delta Q_k$ are computed using the admittance matrix, nodal power balance equation, and~mismatch equations with the data from the last iteration or initial guess. Then, $\vec{\theta}$ and $|\vec{V}|$ are updated by $ \Delta \vec{\theta}$ and $\Delta |\vec{V}|$, respectively. The~algorithm is considered converged when $\| \Delta \vec{P}\|$ and $\| \Delta \vec{Q}\|$ are smaller than a convergence~tolerance. 

It is worth noting that while HHL can solve Equation~\eqref{eq:power-nr} for the normalized solution state $[\Delta \vec{\theta} ^T \,\, \Delta |\vec{V}|^T]^T$ with limited accuracy, the~un-normalized vector could have a smaller norm than the accuracy of HHL. Thus, the~final accuracy of voltage magnitude and phase angles is much higher than the accuracy used in HHL. This situation is similar to iterative refinement in semi-definite optimization in~\cite{mohammadisiahroudi2023quantum}.

\subsubsection{Settings of the Numerical~Experiments}

\begin{figure}[H]
    \includegraphics[width=0.95\linewidth]{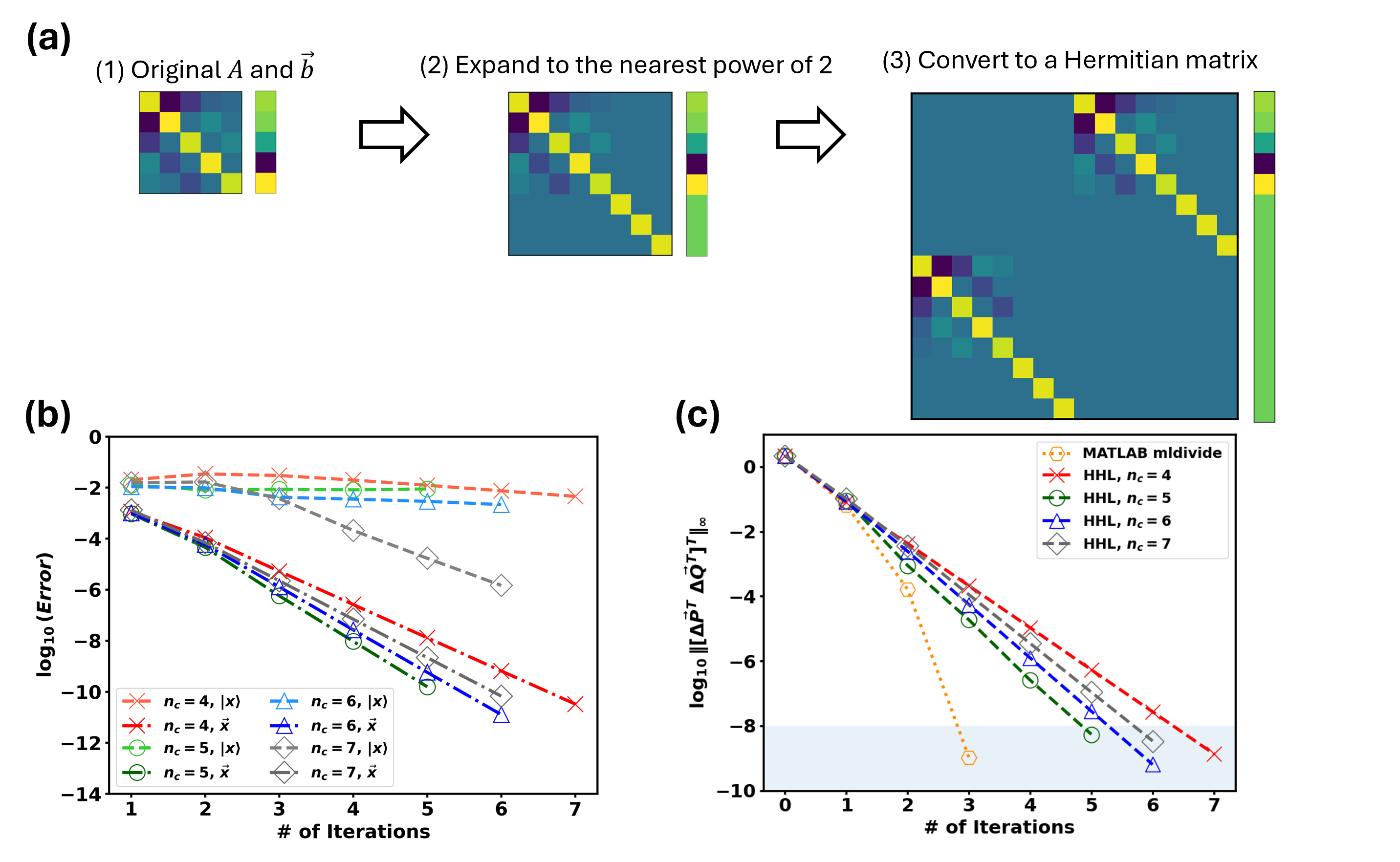}
    \caption{Matrix expansion and error plot for the experiments of the power flow problem. (\textbf{a})~Expanding the coefficient matrices and the RHS vector in Equation~\eqref{eq:power-nr} to fulfill the requirements of the HHL algorithm. (\textbf{b}) {Two types of solution errors in each N-R iteration with various values of $n_c$: the errors of the solution state (labeled ``$\ket{x}$'') in log base 10 , $\|\ket{x} - \ket{x}_{HHL}\|_2$,  and~the errors of the solutions (labeled ``$\vec{x}$'') in log base 10 , $\|\vec{x} - \vec{x}_{HHL}\|_2$.} The symbols $A$, $\vec{b}$, $\vec{x}$, and~$\ket{x}$ refer to the corresponding part in Equation~\eqref{eq:power-nr}. (\textbf{c}) {The convergence performance under different values of $n_c$. The~$y$-axis is the convergence criteria, the~infinity norms of $[\Delta \vec{P}^T \,\, \Delta \vec{Q}^T]^T$, in~log base 10. The value at iteration $0$ represents the norm from the initial guess, and~the gray-shaded area is where the convergence criteria are satisfied.}}
    \label{fig:pf-combo}
\end{figure}

The test case is the four buses and two generators problem in  (\cite{Grainger1994-fo}, p.~377), coded in a MATLAB package called MATPOWER~\cite{matpower}. Based on the framework built in~\cite{zheng2024early}, we incorporate HHL circuits and quantum simulators into the solving process in MATPOWER. The~linear systems of our interest are all $5 \times 5$ systems but not Hermitian. So, the~actual input system is first expanded to $8 \times 8$ so the size of the RHS vector is the power of $2$, and then it is enlarged to $16 \times 16$ following Equation~\eqref{eq:nonher-hhl}. So, we eventually use $4$ qubits to encode the vector $\vec{b}$. This process is illustrated in Figure~\ref{fig:pf-combo}a.

The default value of $n_c$ set by~\cite{nielsen2001quantum} using Equation~\eqref{eq:qishhlnc} is $6$. To~demonstrate how the accuracy of eigenvalues affects an iterative algorithm, we select $n_c$ from $4$ to $7$. With~$4$ clock qubits in QPE and an ancillary qubit required by the HHL algorithm, the~number of qubits in each HHL circuit ranges from $9$ to $12$. The~N-R method converges when
\begin{align*}
    \left\|     \begin{bmatrix}
        \Delta \vec{P} \\ \Delta \vec{Q}
    \end{bmatrix}   \right \|_\infty <~10^{-8}.
\end{align*}
However, because~the linear system formed in an N-R iteration depends on the solution from the previous N-R iteration, the~linear systems at Iteration $j$ with different $n_c$ will differ. Our comparison focuses on the convergence speed and the final solution at the convergence instead of errors at each iteration across different $n_c$.

\subsubsection{Performance~Evaluations}

The sparsity of all tested coefficient matrices is $84.375\%$ after the expansion, with~condition numbers in the range of $[5.950, 5.970]$. The~minimums of the magnitude of eigenvalues are in the range of $[12.263, 12.506]$, and~the maximums are $[73.209, 74.659]$. Figure~\ref{fig:pf-combo}b,c provide illustrative evidence of the use of a less precise linear solver in the iterative method like the N-R method. Although~the N-R method with an HHL subroutine converges slower than a classical linear solver in MATLAB, all methods converge under the same criteria and obtain a similar solution. A~trade-off between convergence speed and complexity of linear system solving exists in our~experiments. 

On the other hand, if~we compare the values of normalized error $\|\ket{x} - \ket{x}_{HHL}\|_2$, when $n_c = 4,5,6,7$, 
using more clock qubits indeed leads to lower error from the HHL algorithm itself. However, increasing $n_c$ does not imply less error on the solution vectors, $\vec{x}_{HHL}$, nor faster convergence by looking at the values of $\|\vec{x} - \vec{x}_{HHL}\|_2$ and $[\Delta \vec{P}^T \,\, \Delta \vec{Q}^T]^T$ in Figure~\ref{fig:pf-combo}. 
The HHL algorithm with $n_c = 5$ gives the fastest convergence, which is smaller than the default value, $6$, from~Equation~\eqref{eq:qishhlnc}.

\subsubsection{Gate Counts and Depths of HHL~Circuits.\label{sec:app-pf-gate}}

Because the circuits from later iterations are in a similar resource demand, we only look at the circuits in the first iteration. The~depths and gate counts of HHL circuits are the same across N-R iterations when $n_c$ is fixed. While HHL with $n_c = 5,6,7$ gives similar convergence speed and accuracy, the~required resources to run the circuits exponentially increase as $n_c$ increases based on Table~\ref{tab:pf-circ}. On~the other hand, although~gate fusion employed in NWQSim does not mitigate these exponential trends, it maintains a constant proportional performance across various HHL circuits: a $79\%$ reduction in gate counts on all tested circuits regardless of the value of $n_c$. 

\begin{table}[H]
    \centering
    \small
    \caption{Depths and gate counts of HHL circuits for power flow problems at Iteration~1.\label{tab:pf-circ}}
    \begin{tabularx}{\textwidth}{cc cc c CC} 
    \toprule
    \boldmath{$n_d$} & \boldmath{$n_c$} & \textbf{Depth}  & \textbf{Gates} &  \textbf{2-Qubit Gates} & \textbf{Gates After Fusion} & \textbf{Reduction from Fusion} \\ \midrule
    4 &4  & 65,824  & 86,262      & 30,651   & 18,060   & $79.06\%$  \\
    4 &5  & 135,986 & 178,180     & 63,315   & 37,283   & $79.08\%$  \\
    4 &6  & 276,308 & 361,980     & 128,631  & 75,717   & $79.08\%$  \\
    4 & 7  & 556,950 & 729,534     & 259,247  & 152,570  & $79.09\%$ \\ \bottomrule
    \end{tabularx}
\end{table}
\unskip

\subsubsection{Resource Estimation in a Fault-Tolerant Scenario~\label{sec:app-pf-re}}

\begin{figure}[H]
    
    \includegraphics[width=0.9\linewidth]{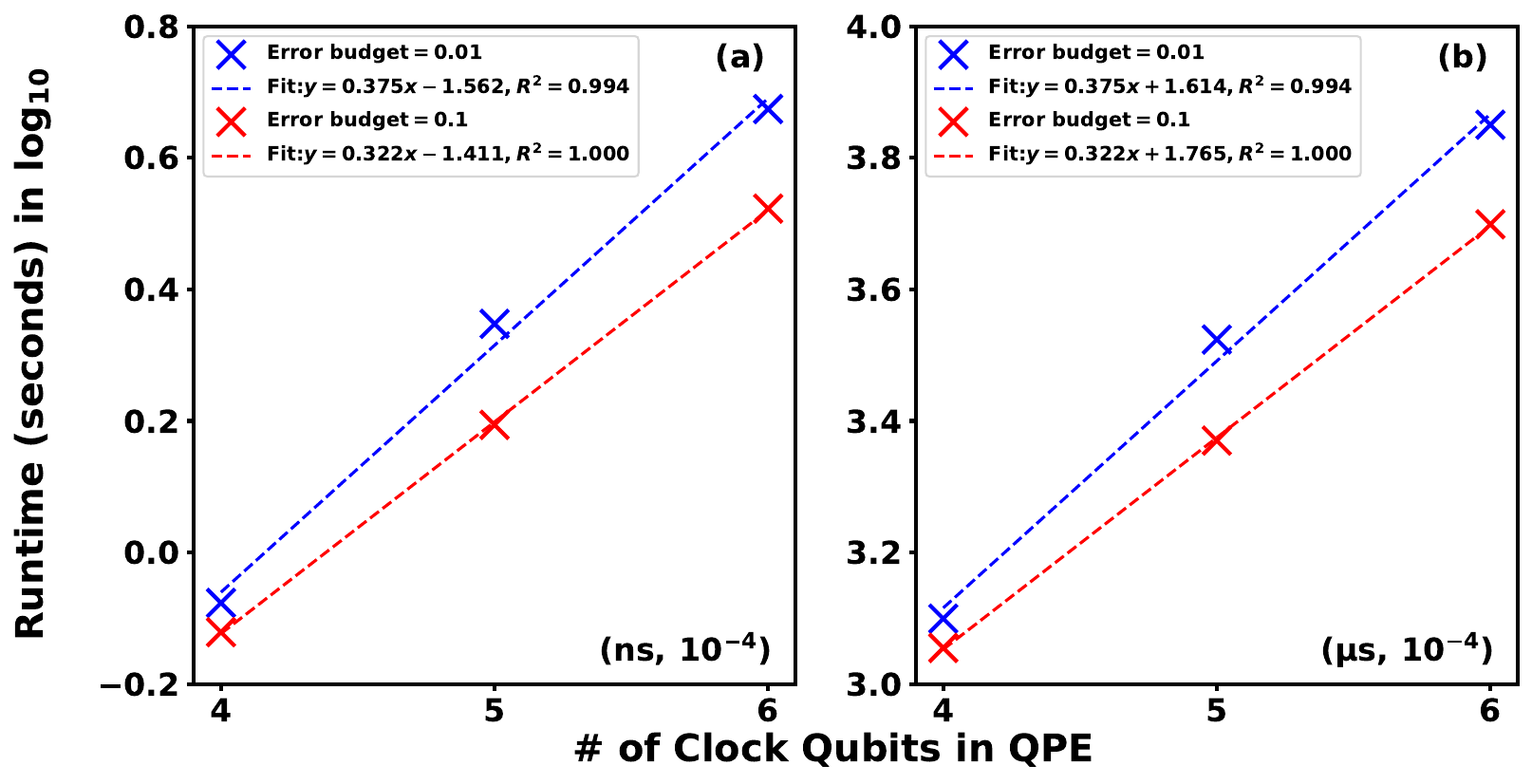}
    \caption{The runtime in seconds as a function of the number of clock qubits in QPE under the qubit parameter set (\textbf{a}) $(\text{ns},~10^{-4})$ and (\textbf{b}) $(\text{$\upmu$s},~10^{-4})$. The~estimated circuits are HHL circuits for power flow~problems.}
    \label{fig:rert-pf}
\end{figure}

\begin{figure}[H]
    
    \includegraphics[width=0.9\linewidth]{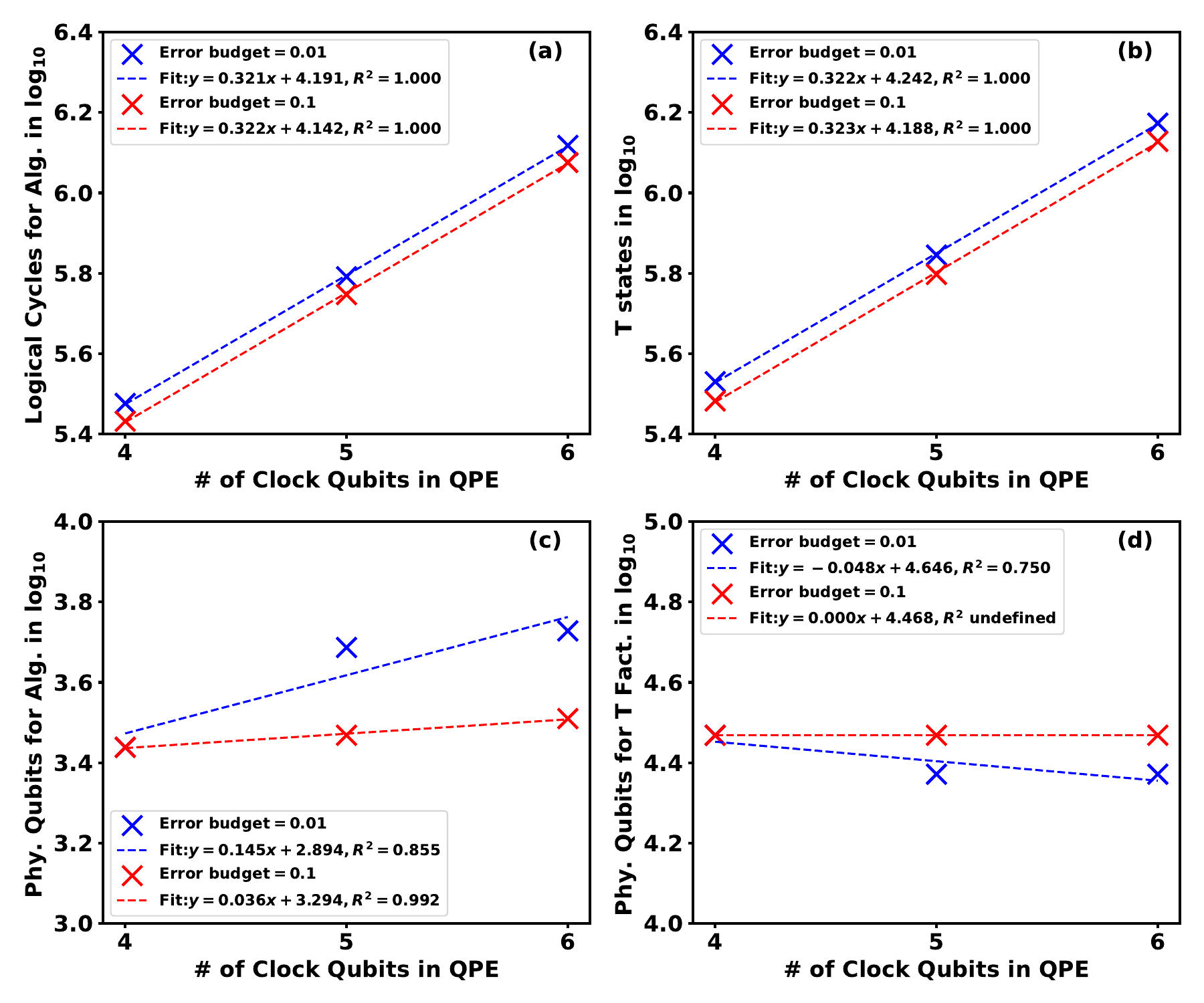}
    \caption{The number of (\textbf{a}) logical cycles for the algorithm, (\textbf{b}) $T$ states, (\textbf{c}) physical qubits for the algorithm after layout, and~(\textbf{d}) physical qubits for the $T$ factories as functions of the number of clock qubits in QPE, respectively. The~estimated circuits are HHL circuits for power flow problems. Both qubit parameter sets $(\text{ns},~10^{-4})$ and $(\text{$\upmu$s},~10^{-4})$ have the same values under the same error budget for all four factors in the~plots.}
    \label{fig:redetail-pf}
\end{figure}

Encoded by the surface code described in Section~\ref{sec:factors} along with a nearest-neighbor connectivity constraint, we estimate the runtime of HHL circuits by the Azure Quantum resource estimator and summarize the data in Figure~\ref{fig:rert-pf}. A~strong and consistent linear correlation between the number of clock qubits in QPE, $n_c$, and~the runtime in log base 10 is displayed across qubit parameter sets and error budgets. Every extra clock qubit brings $10^{0.322} \approx 2.099$ times longer runtime when the error budget is $0.1$ and $10^{0.375} \approx 2.371$ times longer when the error budget is $0.01$. This multiplier shows an increasing trend when the error budget decreases. Similar correlations are also demonstrated in Figure~\ref{fig:redetail-pf}a,b when we further investigate how $n_c$ affects the number of logical cycles for the circuit and the number of $T$ states. Generally, the~exponential dependencies of runtime, number of logical cycles, and~number of $T$ states on $n_c$ match the relationship between the number of gates in HHL circuits and $n_c$. Note that the slopes of the fitted line in Figure~\ref{fig:redetail-pf}a,b are not sensitive to error budgets, different from the behavior in Figure~\ref{fig:rert-pf}. Error budgets affect the constant multiplier of the growth of logical cycles and the number of $T$ states~more.

Table~\ref{tab:pf-qubits} summarizes the other factors of our interest. Those factors have the same values in $(\text{ns},~10^{-4})$ and $(\text{$\upmu$s},~10^{-4})$ settings. Note that there is a dramatic fall in the number of physical qubits when the error budget is 0.01 and $n_c$ raises from 4 to 5. Combined with Figure~\ref{fig:redetail-pf}c,d, this reduction comes from a large drop in the number of physical qubits spent on $T$ factories, a~dominant demand on physical qubits instead of the quantum algorithm itself. The~circuit requires $15$ $T$ factories when the error budget is $0.01$ and $n_c = 4$, but~this number is reduced to $12$ when $n_c = 5$. Recall the definition of the number of $T$ factories in Table~\ref{tab:factors-qec}, based on the fitted coefficients in Figures~\ref{fig:rert-pf} and~\ref{fig:redetail-pf}; we can see while the increase in $n_c$ from $4$ to $5$ leads to $10^{0.322}$ times more $T$ states, the~runtime becomes $10^{0.375}$ times larger. Since $T$ factory duration and $T$ states per factory are kept constant, the~faster-growing runtime reduces the number of $T$ factories required, thus decreasing the overall number of physical qubits required. This phenomenon does not occur when the error budget is $0.1$ because the growth of runtime and $T$-state count are at the same~speed.

\begin{table}[H]
    \centering
    \caption{Factors of interest for fault-tolerant HHL circuits in power flow~problems.\label{tab:pf-qubits}}
    \footnotesize
    \begin{tabularx}{\textwidth}{c c CCC c} 
    \toprule
    \textbf{Error}     & \multirow{2}{*}{\boldmath{$n_c$}} & \textbf{Physical Qubits}  & \textbf{Logical Qubits}            & \textbf{Min. Logical Qubit}   & \textbf{Min.} \boldmath{$T$} \textbf{State}  \\ 
    \textbf{Budget}    &                         & \textbf{After Layout}    & \textbf{Pre- and After Layout} & \textbf{Error Rate}     &  \textbf{Error Rate} \\ 
    \midrule
    \multirow{3}{*}{0.01} & 4  & 32,144              & 9 to 28           & 3.977 $\times~10^{-10}$     & 9.831 $\times~10^{-9}$      \\ 
                          & 5  & 28,380              & 10 to 30           & 1.797 $\times~10^{-10}$     & 4.762 $\times~10^{-9}$      \\
                          & 6  & 28,866              & 11 to 33           & 7.700 $\times~10^{-11}$     & 2.236 $\times~10^{-9}$      \\ \midrule
    \multirow{3}{*}{0.1}  & 4  & 32,144              & 9 to 28           & 4.406 $\times~10^{-9}$     & 1.098 $\times~10^{-7}$      \\
                          & 5  & 32,340              & 10 to 30           & 1.990 $\times~10^{-9}$     & 5.319 $\times~10^{-8}$      \\
                          & 6  & 32,634              & 11 to 33           & 8.487 $\times~10^{-10}$     & 2.483 $\times~10^{-8}$     \\ \bottomrule
    \end{tabularx}
\end{table}

\vspace{-6pt}\vspace{-6pt}


\subsection{Heat Transfer Problem in Climate~Projection}\label{sec:heat}

{Linear solvers are deeply embedded in linear or non-linear differential equation solving through numerical methods such as the Carleman linearization and the finite difference method~\cite{li2023potential}}.
Such methods discretize the domain of the problems into grids, and~the dimension of the formed linear system scales as the size of discretization. The~number of grid points scales polynomially with system size, while the demands for solving such differential equations (DEs) are ubiquitous in science and engineering. Due to the exponential speedup in problem dimension, the~combination of quantum linear solvers and these numerical methods has become an attractive direction~\cite{montanaro2016quantum, childs2021high, saha2022advancing, li2023potential}. For~example, accurate climate projection, one of the most scientifically challenging and socially urgent problems, is cursed by high dimension and could be revolutionized by quantum computing. In~this section, we explore the application of a quantum linear system solver to the heat transfer equation that is important for atmospheric processes related to climate~projection. 

\subsubsection{Settings of the Numerical~Experiments}

In this section, we examine the two-dimensional (2-D) heat diffusion equation in~\cite{li2019cloud}
\begin{align}
    \frac{\partial T}{\partial t} = D \nabla^2 T + F \label{eq:heat-de}
\end{align}
where $T$ represents the temperature at a given 2-D point and time, $D$ is the heat transfer coefficient, and~$F$ is the forcing term consisting of arbitrary boundary and initial conditions. \Eq{eq:heat-de} is a linear partial differential equation. We discretize \Eq{eq:heat-de} in space and time into a system of ordinary differential equations using the finite difference method,
\begin{equation}\label{eq:HHL}
    A T = F, 
\end{equation}
where $A$ is the resulting coefficient matrix.
Take the square lattices with a lateral size of 
three grid points and five grid points, and the~resulting dimension of $A$ is $9 \times 9$ and $25 \times 25$, respectively. Such configurations require $4$ qubits and $5$ qubits to represent the RHS vectors ($F$ term in \Eq{eq:HHL}) in both linear systems, respectively. Let $A^{(heat, l)}$ be the coefficient matrix generated from $l$ number of grid points; the~entry values are
\begin{align*}
    A^{(heat, l)}_{pq} = \begin{cases} 
    1+4r, & p = q \\
    -r & p = q \pm 1 \text{ or } \text{ or } p = q \pm~l \\
    0, &\text{otherwise}
    \end{cases}
\end{align*}
where $p$ and $q$ denote the index of the entries of $A$, and $r$ is $0.00016$ in the three-point case and $0.00064$ in the five-point~case.

\subsubsection{Performance and Resource~Evaluations}

The coefficient matrices are Hermitian by design, so we only need to expand the dimension to the nearest power of $2$, i.e.,~$16$ and $32$. After~dimension expansion, the~coefficient matrices have sparsity $82.81\%$ and $88.28\%$, respectively. Both matrices have condition number $1$, and~all of their eigenvalues are around $1$.

\begin{figure}[H]
    
    \includegraphics[width=0.7\linewidth]{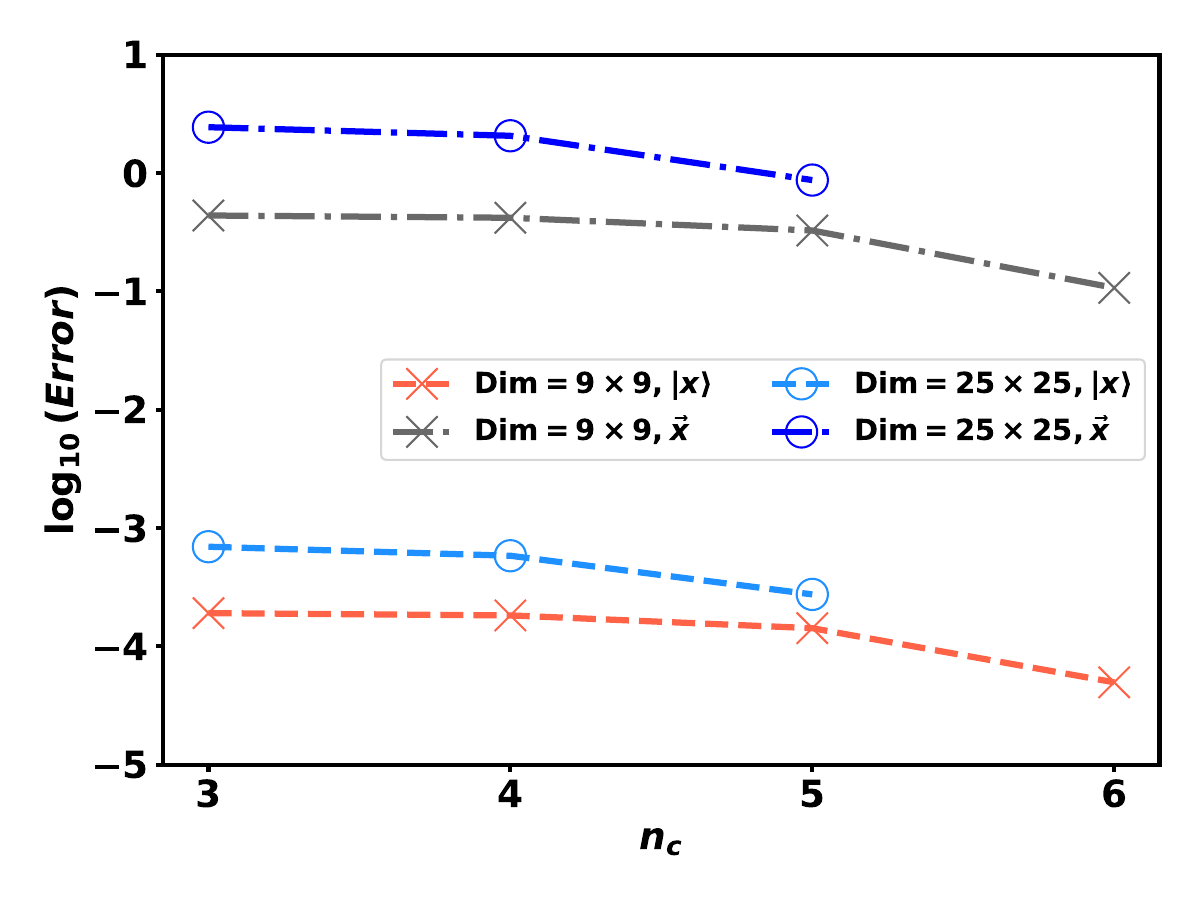}
    \caption{ The errors of the solution states in log base 10 (labeled as ``$\ket{x}$''), $\|\ket{x} - \ket{x}_{HHL}\|_2$, and~the errors of the solution vectors in log base 10 (labeled as ``$\vec{x}$''), $\|\vec{x} - \vec{x}_{HHL}\|_2$, are presented as functions of $n_c$ for two different numbers of grid points. The~symbols $\vec{x}$ and $\ket{x}$, respectively, refer to the solution vector and the normalized solution~vector.}
    \label{fig:atoms-err}
\end{figure}

When $n_d = 4$, gate counts in Tables~\ref{tab:pf-circ} and~\ref{tab:atoms-circ} have almost the same numbers of circuit depths and gate counts. However, if~we compare across different $n_d$ in Table~\ref{tab:atoms-circ}, significant increases appear in depths and all gate counts. This situation reflects one of Aaronson's concerns in~\cite{Aaronson2015} about the efficiency and the cost of data reading in quantum linear solvers. Furthermore, similar to the scenario in Section~\ref{sec:power-grid}, the~incremental of $n_c$, despite being very costly, has a limited contribution towards reducing errors, as~shown in Figure~\ref{fig:atoms-err}.

\begin{table}[H]
    \centering
    \small
    \caption{Depths and gate counts of HHL circuits for heat transfer~problems.\label{tab:atoms-circ}}
   
\begin{adjustwidth}{-\extralength}{0cm}
 \begin{tabularx}{\fulllength}{cccc c CCC} \toprule
    \textbf{Dim.} & \boldmath{$n_d$} & \boldmath{$n_c$} & \textbf{Depth}  & \boldmath{$\#$} \textbf{of Gates} & \boldmath{$\#$} \textbf{of 2-Qubit Gates} & \boldmath{$\#$} \textbf{of Gates After Fusion} & \textbf{Reduction from Fusion} \\ \midrule
    $9 \times 9$        & 4  & 3  & 30,742
   & 40,290      & 14,315   & 8445          & 79.04\%        \\
    $9 \times 9$        & 4  & 4  & 65,824   & 86,262      & 30,651   & 18,061         & 79.06\%        \\
    $9 \times 9$        & 4  & 5  & 135,986  & 178,180     & 63,315   & 37,284         & 79.08\%        \\
    $9 \times 9$        & 4  & 6  & 276,308  & 361,980     & 128,631  & 75,718         & 79.08\%        \\ \midrule
    $25 \times 25$       & 5  & 3  & 133,966  & 175,253     & 62,546   & 37,147         & 78.80\%        \\
    $25 \times 25$       & 5  & 4  & 287,134  & 375,643     & 134,046  & 79,547         & 78.82\%        \\
    $25 \times 25$       & 5  & 5  & 593,948  & 777,069     & 277,230  & 164,338        & 78.85\%        \\ \bottomrule
    \end{tabularx}
\end{adjustwidth}
\end{table}

\subsubsection{Resource Estimation in a Fault-Tolerant~Scenario}

\begin{figure}[H]
    
    \includegraphics[width=0.9\linewidth]{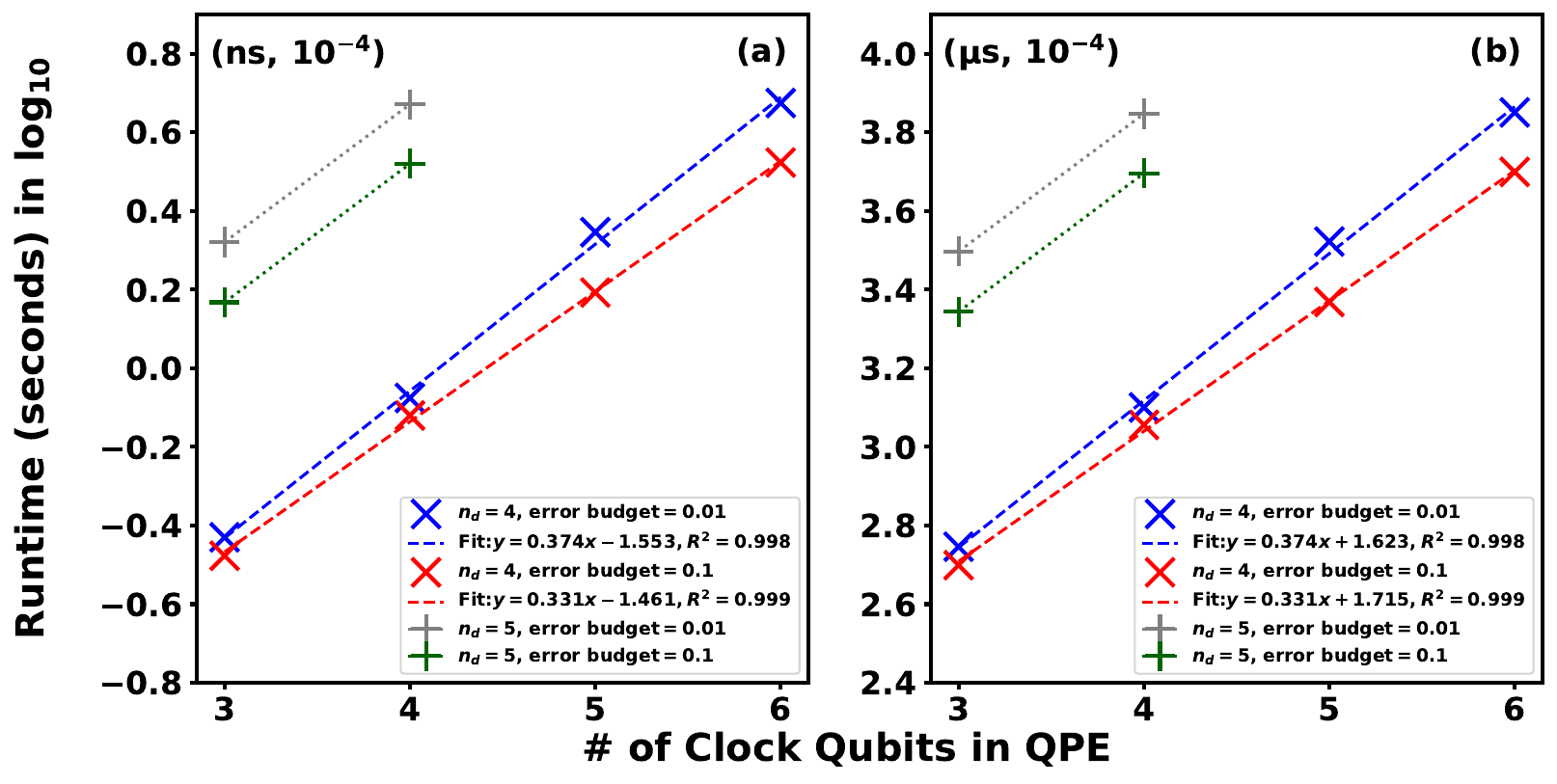}
    \caption{The runtime in seconds as a function of the number of clock qubits in QPE under the qubit parameter set (\textbf{a}) $(\text{ns},~10^{-4})$ and (\textbf{b}) $(\text{$\upmu$s},~10^{-4})$. The~estimated circuits are HHL circuits for the heat transfer~problem.}
    \label{fig:rert-atoms}
\end{figure}

\begin{figure}[H]
    
    \includegraphics[width=0.9\linewidth]{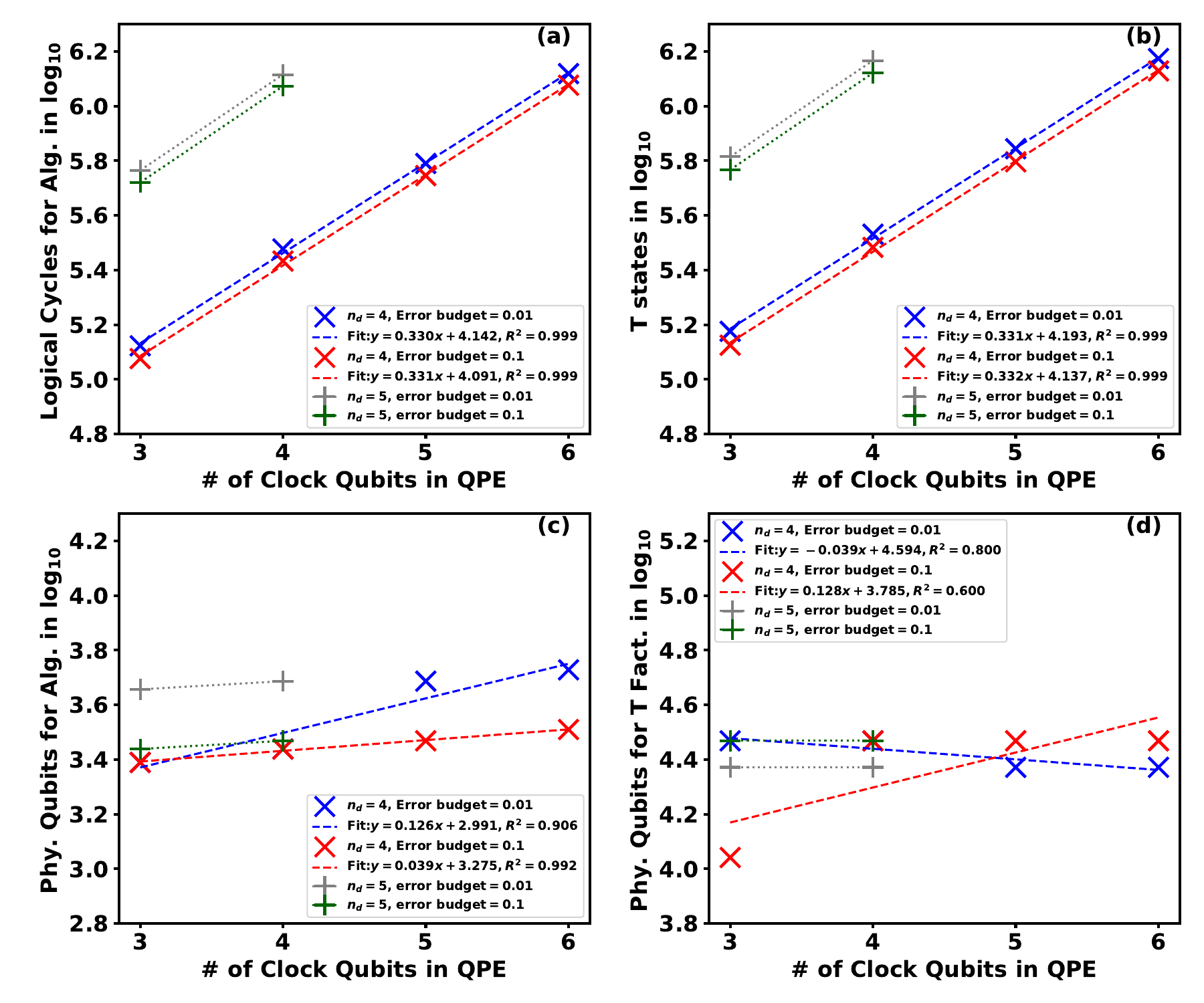}
    \caption{The number of (\textbf{a}) logical cycles for the algorithm, (\textbf{b}) $T$ states, (\textbf{c}) physical qubits for the algorithm after layout, and~(\textbf{d}) physical qubits for the $T$ factories as functions of the number of clock qubits in QPE, respectively. The~estimated circuits are HHL circuits for the heat transfer problem. Both qubit parameter sets $(\text{ns},~10^{-4})$ and $(\text{$\upmu$s},~10^{-4})$ have the same values under the same error budget for all four factors in the~plots.}
    \label{fig:redetail-atoms}
\end{figure}

Most of the observations from Figures~\ref{fig:rert-atoms} and~\ref{fig:redetail-atoms} and Table~\ref{tab:atoms-qubits} for both problem sizes are analogous to the findings in Section~\ref{sec:app-pf-re}, including the numerical values of the fitted-line coefficients related to runtime, logical cycles, and the number of $T$ states. The~significant influence brought by deeper data-loading modules for the five-point problem is parallel shifts on longer runtime, more logical cycles, more $T$ states, and~more strict requirements on the logical qubit error rate and $T$ state error rate. More data-loading qubits do not affect the growth speed of the logical cycle and the number of $ T $ states. Due to the limitation of computational time in Azure Quantum cloud service, we cannot collect more data points to understand this correlation better. However, from~a theoretical perspective, this is expected because the QPE costs of HHL circuits are the same with the same $n_c$ in the power flow and heat transfer~problems.

\vspace{-6pt}\vspace{-3pt}
\begin{table}[H]
    \centering
    \caption{Factors of interest for fault-tolerant HHL circuits in heat transfer~problems.\label{tab:atoms-qubits}}
    \small
    \begin{tabularx}{\textwidth}{cccccc} \toprule
    \textbf{Error}     & \multirow{2}{*}{\boldmath{$(n_d,\, n_c)$}} & \textbf{Physical Qubits}  & \textbf{Logical Qubits }           & \textbf{Min. Logical Qubit}   & \textbf{Min.} \boldmath{$T$} \textbf{State}  \\ 
    \textbf{Budget}    &                         & \textbf{After Layout}    & \textbf{Pre- and After Layout}     & \textbf{Error Rate}     &  \textbf{Error Rate} \\ 
    \midrule
\multirow{6}{*}{0.01}         & (4, 3)
                      & 31,850                                  & 8 to 25                               & 1.01 $\times~10^{-9}$                          & 2.22 $\times~10^{-8}$                           \\
                              & (4, 4)                      & 32,144                                  & 9 to 28                               & 3.97 $\times~10^{-10}$                          & 9.81 $\times~10^{-9}$                           \\
                              & (4, 5)                      & 28,380                                  & 10 to 30                               & 1.80 $\times~10^{-10}$                          & 4.77 $\times~10^{-9}$                           \\
                              & (4, 6)                      & 28,866                                  & 11 to 33                               & 7.69 $\times~10^{-11}$                          & 2.23 $\times~10^{-9}$                           \\
                              & (5, 3)                      & 28,056                                  & 9 to 28                               & 2.05 $\times~10^{-10}$                          & 5.11 $\times~10^{-9}$                           \\
                              & (5, 4)                      & 28,380                                  & 10 to 30                               & 8.53 $\times~10^{-11}$                          & 2.27 $\times~10^{-9}$                           \\  \midrule
\multirow{6}{*}{0.1}          & (4, 3)                      & 13450                                  & 8 to 25                               & 1.12 $\times~10^{-8}$                          & 2.50 $\times~10^{-7}$                           \\
                              & (4, 4)                      & 32,144                                  & 9 to 28                               & 4.40 $\times~10^{-9}$                          & 1.10 $\times~10^{-7}$                           \\
                              & (4, 5)                      & 32,340                                  & 10 to 30                               & 2.00 $\times~10^{-9}$                          & 5.33 $\times~10^{-8}$                           \\
                              & (4, 6)                      & 32,634                                  & 11 to 33                               & 8.47 $\times~10^{-10}$                          & 2.48 $\times~10^{-8}$                           \\
                              & (5, 3)                      & 32,144                                  & 9 to 28                               & 2.27 $\times~10^{-9}$                          & 5.70 $\times~10^{-8}$                           \\
                              & (5, 4)                      & 32,340                                  & 10 to 30                               & 9.39 $\times~10^{-10}$                          & 2.52 $\times~10^{-8}$  \\ \bottomrule                        
    \end{tabularx}
\end{table}

\section{Discussion}\label{sec:discussion}

This paper evaluates and analyzes the performance and resources required for the HHL algorithm in various scientific and engineering problems. There are still multiple points we need to address in future work. The~foremost limitation in this work is the data-loading module in the HHL circuit generation. While the data-loading algorithm in~\cite{anedumla_quantum_linear_solvers} can encode an arbitrary vector into a quantum circuit, the~circuit depth of this module is exponential in the number of qubits. Thus, this first part of the circuit severely damages the potential quantum speedup from HHL. We mitigate this drawback by comparing the outcomes from problems of different sizes to isolate the influence of the data-loading module. An~important future direction is incorporating an efficient data-loading scheme into our analysis framework, like block-encoding in~\cite{casares2020quantum}. A~different data-loading method could have a different accuracy, so it is necessary to investigate how data-loading accuracy and condition number of coefficient matrices collectively affect the solution accuracy. This future direction illustrates the second drawback of this study. That is, our tested coefficient matrices are all well-conditioned. Because~our experiments do not utilize randomly generated test cases, we have less control over the matrix properties, including condition number and sparsity. 
A potential source of ill-conditioned test cases is the methods that naturally have ill-conditioned matrices, such as the Newton systems produced by the interior-point method in optimization problems~\cite{mohammadisiahroudi2023quantum}. 
Thus, to~solve those systems, iterative refinement with the HHL algorithm~\cite{mohammadisiahroudi2022efficient} and a variant of the HHL algorithm in~\cite{clader2013preconditioned}, accompanied by the sparse approximate inverse preconditioner, is in our outlook.
Limited by the single-job running time in the Azure Quantum cloud server, we cannot process large HHL circuits, mainly limited by the number of gates. This restricts the number of clock qubits in the QPE and the number of data points in each plot in Section~\ref{sec:applications}. This is why we only discuss the correlations whose coefficients of determination are almost $1$. In~future studies, we will dismantle the whole HHL circuit into different modules and evaluate the resource cost separately. 

Some additional research can be conducted to further enhance our understanding of the application of quantum algorithms in scientific problems. An~important direction is understanding the implication of various noise models on the HHL algorithm. We plan to conduct those experiments with the high-precision noise simulator in~\cite{li2024tanq}. We can also include the quantum algorithms that address similar scientific applications into our resource analysis framework, such as quantum differential equation solvers in~\cite{PhysRevLett.131.150603,an2023quantum} and quantum optimizers in~\cite{QHD, wiebeopt}.

\section{Conclusions}\label{sec:conclusion}

In this paper, we investigate the practical applications and scalability of the HHL algorithm in solving quantum linear systems associated with scientific problems like power grids and heat transfer problems. Through the NWQSim package on high-performance computing platforms, we highlight the benefits of the utilization of low-accuracy QPE in HHL for both iterative and non-iterative methods in practice: low-accuracy QPE can exponentially reduce the gate counts and circuit depth in an HHL circuit, while keeping the same solution accuracy in iterative methods like the Newton--Raphson method and maintaining a similar level of accuracy in a non-iterative method like the finite difference~method. 

Furthermore, with the Azure Quantum resource estimator, we evaluate the resource requirements of HHL circuits in our experiments under two settings that simulate superconducting and trapped-ion qubits. The~correlations between QEC-related criteria and the input HHL circuits have been thoroughly studied. The~runtime, number of logical cycles, and~number of $T$ states have exponential dependencies on the number of clock qubits in QPE. However, this relation is not necessarily inherited by the number of physical qubits demanded. 
In our experiments, we find that even as $n_c$ increases and the error budget reduces, it is possible that $T$ factory demand also decreases.
More specifically, if~the runtime growth is faster than the required amount of $T$ states, the~circuit needs fewer $T$ factories and thus fewer physical qubits to prepare $T$ factories. Since the growth of runtime is sensitive to the error budget, it is possible to reduce the physical qubit requirement if a low error budget is achievable on early fault-tolerant quantum devices.

Our study provides pivotal insights into the operational requirements of quantum linear system algorithms, paving the way for further empirical studies. We propose future research on the applications of quantum linear system solvers and iterative refinement on high-fidelity quantum computers for small-scale experiments. For~large-scale experiments, we suggest using noise-modeled simulators on high-performance platforms. In~the context of QEC and early fault-tolerant quantum computing, we believe it is crucial to focus on controlling the resource cost of $T$ factories by considering the runtime and error budget. These research directions hold promise for bridging the gap between theoretical potential and practical usability in quantum~computing.
All the code in this research will be hosted in a public repository (\url{https://github.com/pnnl/nwqlib}).

\vspace{6pt} 

\authorcontributions{Conceptualization, M.Z., C.L., S.S., X.L., J.M., Y.C., and~A.L.; methodology, M.Z., C.L., S.S., and~A.L.; software, M.Z. and~A.L.; validation, S.S., X.L., J.M., Y.C., and~A.L.; formal analysis, M.Z., and C.L.; investigation, M.Z.; resources, X.L., J.M., Y.C., and~A.L.; data curation, M.Z., X.L., J.M., and~Y.C.; writing---original draft preparation, M.Z., C.L., S.S., X.L., J.M., Y.C., and~A.L.; writing---review and editing, M.Z. and~A.L.; visualization, M.Z.; supervision, S.S., J.M., Y.C. and~A.L.; project administration, A.L.; funding acquisition, A.L. All authors have read and agreed to the published version of the manuscript.}

\funding{This research was supported by the Quantum Algorithms and Architecture for Domain Science Initiative (QuAADS), under~the Laboratory Directed Research and Development (LDRD) Program at Pacific Northwest National Laboratory (PNNL). This research used resources of the Oak Ridge Leadership Computing Facility at the Oak Ridge National Laboratory, which is supported by the Office of Science of the U.S. Department of Energy under Contract
No. DE-AC05-00OR22725. This research used resources of the National Energy Research Scientific Computing Center (NERSC), a~U.S. Department of Energy Office of Science User Facility located at Lawrence Berkeley National Laboratory, operated under Contract No. DE-AC02-05CH11231. PNNL is a multi-program national laboratory operated for the U.S. Department of Energy (DOE) by Battelle Memorial Institute under Contract No. DE-AC05-76RL01830.}

\dataavailability{The original contributions presented in this study are included in the article.
Further inquiries can be directed to the corresponding author(s).} 


\conflictsofinterest{The authors declare no conflicts of~interest.}

\begin{adjustwidth}{-\extralength}{0cm}

\reftitle{References}



%


\PublishersNote{}
\end{adjustwidth}
\end{document}